\documentclass[useAMS,usenatbib]{mn2e}

\usepackage{graphicx} \usepackage{natbib}

\newcommand{\logg}{\mbox{$\log g$}}

\def\m2s2{\hbox{\,m$^{2}$\,s$^{-2}$}} %m2.s -2
\def\Msun{\hbox{$\mathrm{M}_{\odot}$}} %Msun
\def\Rsun{\hbox{$\mathrm{R}_{\odot}$}} 
\def\Mjup{\hbox{$\mathrm{M}_{\rm Jup}$}}
\def\Rjup{\hbox{$\mathrm{R}_{\rm Jup}$}}

\def\1s{$1\,\sigma$} 
 
\def \t0{T$_0$}

\usepackage{times}
\title[TTVs in WASP-10b induced by stellar activity?]{Transit timing variations in WASP-10b induced by stellar activity?}
\author[S.C.C. Barros et al.]{S. C.  C. Barros $^{1}$\thanks{E-mail:susana.barros@oamp.fr}, G. Bou\'e $^{2,3,4}$, N. P. Gibson $^{5}$,  D. L. Pollacco$^{6}$,  A. Santerne $^{1}$, \and F. P. Keenan $^{7}$, I. Skillen$^{8}$, R. A. Street$^{9}$\\
  $^{1}$ Aix Marseille Universit\'e, CNRS, LAM (Laboratoire d'Astrophysique de Marseille) UMR 7326, 13388, Marseille, France \\
  $^{2}$ Centro de Astrof\'isica, Universidade do Porto, Rua das Estrelas, 4150-762 Porto, Portugal \\
   $^{3}$ Astronomie et Syst\`emes Dynamiques, IMCCE-CNRS UMR8028, Observatoire de Paris, UPMC, 77 Av. Denfert-Rochereau, 75014 Paris, France \\
    $^{4}$          Department of Astronomy and Astrophysics, University of Chicago, 5640 South Ellis Avenue, Chicago, IL 60637, USA \\
  $^{5}$European Southern Observatory, Karl-Schwarzschild-Str. 2, 85748 Garching bei Mu ̈nchen, Germany\\
   $^{6}$ Department of Physics, University of Warwick, Coventry CV4 7AL, UK \\
  $^{7}$Astrophysics Research Centre, School of Mathematics and  Physics, Queen's University Belfast, University Road, Belfast, BT7  1NN, UK\\
   $^{8}$Isaac Newton Group of Telescopes, Apartado de Correos 321, E-38700 Santa Cruz de la Palma, Tenerife,  Spain\\
 $^{9}$Las Cumbres Observatory, 6740 Cortona Drive Suite 102, Goleta, CA 93117, USA \\ }
 
\begin{document}

\date{Accepted 2013 January 16,  Received 2013 January 15, in original form 2012 November 14}

\pagerange{\pageref{firstpage}--\pageref{lastpage}} \pubyear{2002}

\maketitle

\label{firstpage}

\begin{abstract}
The hot-Jupiter WASP-10b was reported by \citet{Maciejewski2011,Maciejewski2011b} to show transit timing variations (TTV) with an amplitude of $\sim 3.5$ minutes.
These authors proposed that the observed TTVs were caused by a 0.1  \Mjup\  perturbing companion  with an orbital period of $\sim 5.23\,$d, and hence, close to the outer 5:3 mean motion resonance with WASP-10b.
 To test this scenario, we present eight new transit light curves of WASP-10b obtained with the Faulkes Telescope North and the Liverpool Telescope. The new light curves, together with 22 previously published ones, were modelled with a Markov-Chain Monte-Carlo transit fitting code.

Transit depth differences reported for WASP-10b are thought to be due to star spot induced brightness modulation of the host star. Assuming the star is brighter at the activity minimum, we favour a small planetary radius. We find  $ \mathrm{R}_p  =1.039^{+0.043}_{-0.049}\, \Rjup$ in agreement with \citet{Johnson2009} and  \citet{Maciejewski2011b}. \citet{Maciejewski2011} and \citet{Husnoo2012} find no evidence for a significant eccentricity in this system. We present consistent system parameters for a circular orbit and refine the orbital ephemeris of WASP-10b.

Our homogeneously derived transit times do not support the previous claimed TTV signal, which was strongly dependent on 2 previously published transits that have been incorrectly normalised.
Nevertheless, a linear ephemeris is not a statistically good fit to the transit times of WASP-10b. We show that the observed transit time variations are due to spot occultation features or systematics. We discuss and exemplify the effects of occultation spot features in the measured transit times and show that despite spot occultation during egress and ingress being difficult to distinguish in the transit light curves, they have a significant effect in the measured transit times. We conclude that if we account for spot features, the transit times of WASP-10 are consistent with a linear ephemeris with the exception of one transit (epoch 143) which is a partial transit. Therefore, there is currently no evidence for the existence of a companion to WASP-10b. Our results support the lack of TTVs of hot-Jupiters reported for the Kepler sample.

\end{abstract}


\section{Introduction}

The discovery of WASP-10b, a $~ 3 $ \Mjup\  hot-Jupiter planet in a 3.09 day orbit, was reported by \citet{Christian2009}. 
Its host star is a K5V type with $T_{eff} = 4675 \pm 100\,$K,  {[M/H]} $= 0.03 \pm 0.2$ \citep{Christian2009} and is relatively young, with an age of $270\pm 80\,$ Myr \citep{Maciejewski2011}.
Using a high signal-to-noise transit light curve, \citet{Johnson2009} updated the stellar density and re-derived the stellar mass  to $M_*= 0.75 \pm 0.04\,$\Msun\ and radius to  $R_*=0.698 \pm 0.012 \,$\Rsun .
There has been some discussion about the planetary size, with \citet{Christian2009}, \citet{Dittmann2010} and \citet{Krejcova2010} obtaining a planetary radius of $ 1.28 \pm 0.09 $ \Rjup\, while \citet{Johnson2009} derived a smaller radius of $ 1.08 \pm 0.02 $ \Rjup\ . 
\citet{Maciejewski2011b} argue that the discrepancy in previous results was due star spot induced brightness variability.
In fact, WASP-10 has been shown to have stellar variability due to the rotation modulation of spots with a period of  $11.91\pm 0.05\,$ d \citep{Smith2009} and semi-amplitude up to $10.1\,$mmag. Star spots reduce the effective stellar disc area leading to an underestimation of the stellar radius and a larger transit depth. This variation of transit depth with spot coverage explains the differences in the measured planetary radius (e.g. \citealt{Czesla2009}). Using a transit taken close to maximum brightness that corresponds to a minimum spot coverage, \citet{Maciejewski2011b} obtained a maximum planetary radius of $1.03^{+0.07}_{-0.03}$ \Rjup.

Interestingly, \citet{Maciejewski2011} reported a periodic modulation of transit timing variations (TTVs) for WASP-10 with an amplitude of $\sim 3.5 \pm 1.0$ minutes. They suggested a 0.1  \Mjup\  planet companion in the outer 5:3 mean motion resonant orbit as the cause of the measured TTVs. Later, \citet{Maciejewski2011b} confirmed the TTV solution and  showed that only spots located close to the stellar limb would affect the transit times but they would also result in transit duration variations. Since no transit duration variations were found in their sample, the authors argued that spot occultation features did not significantly affect the measured transit times of WASP-10.

The possibility of finding additional planets by measuring their effect on the central transit times of known transiting planet systems was proposed by \citet{Holman2005} and \citet{Agol2005}. They showed that for systems near mean-motion resonances this method can be sensitive to planets less massive than the Earth. This has led many groups to search for TTVs from the ground. However, most of the searches resulted in only upper limits on the mass of possible companions (e. g. \citealt{MillerRicci2008,Bean2009,Gibson2009, Gibson2010}). The few initial reports of TTVs were unconfirmed later. For example, \citet{Diaz2008} found TTVs in OGLE-111b which were unconfirmed in later studies  \citep{Adams2010,Hoyer2011}. At present, only three hot-Jupiter transiting systems show TTVs that were measured from ground-based data: WASP-3 \citep{Maciejewski2010}, WASP-10 \citep{Maciejewski2011}, and WASP-5 \citep{Fukui2011}. Meanwhile, Kepler data has allowed the discovery of many TTV systems (e.g. Kepler-9 \citep{Holman2010} and Kepler-11 \citep{Lissauer2011} ).
However, a recent TTV study of the Kepler sample revealed the lack of TTVs in hot-Jupiters, contrasting with the TTVs found for longer-period Jupiters or hot-Neptunes  \citep{Steffen2012}. The authors suggested that this implied a different formation and/or evolution history for hot-Jupiters. Therefore, it is very important to confirm and understand the hot-Jupiter TTV detections that have been reported from ground-based data.

To confirm and possibly characterise WASP-10b's companion we obtained new high precision transit light curves of WASP-10b. Six transit observations of WASP-10b were obtained with the RISE fast CCD camera mounted on the Liverpool Telescope and two more with the Merope camera on the Faulkes Telescope North.
We combined the new observations with previously published 22 light curves of WASP-10 to homogeneously derive the transit times.

The new observations are described in Section 2. In Section 3, we discuss our transit model.
We present the updated parameters of the system in Section
4 and in Section 5 discuss our results and show possible causes of spurious TTVs. Finally, we present our conclusions in Section 6.

\section{Observations}

We present further high precision transit observations of WASP-10b taken with Faulkes Telescope North (FTN) and with the Liverpool Telescope (LT).

Two full transits of WASP-10b were observed on the 2008-09-15 and 2008-09-19 with the Merope camera on the FTN using the SDSS-i′ filter. The Merope camera consists of a e2V, $2048 \times  2048$ pixel-CCD and has a field of view of 4.7' $\times$ 4.7'. Aperture photometry was performed with the DAOPHOT package \citep{Stetson1987} within the IRAF4 environment. The differential photometry was performed relative to at least 5 comparison stars that were within the FTN field-of-view. 

Another six transits were observed with the fast CCD camera RISE \citep{rise2008,Gibson2008} mounted on the 2.0m Liverpool
Telescope on La Palma, Canary Islands. RISE has a wideband filter covering
$\sim 500$--$700\,$nm which corresponds approximately to V+R. The pixel scale is 0.54 arcsec/pixel resulting in a 9.4'
$\times$ 9.4' field-of-view. Thanks to its frame transfer CCD, RISE has a deadtime of only 35 ms for
exposures longer than 1 second. To decrease systematic noise due to poor guiding \citep{Barros2011b} the telescope was defocussed to -0.6mm and we used an exposure time of 9 seconds. However, in one of the observations (2010-11-10) we experimented with a higher defocussing of -1.0mm and longer exposure time of 34 seconds. The RISE data were reduced using the \textsc{ultracam} pipeline
\citep{Ultracam} following the same procedure as for WASP-21b \citep{Barros2011b}. Details of the observations are given in Table~\ref{obser}.

\begin{figure*}
  \centering
 \includegraphics[width=1.9\columnwidth]{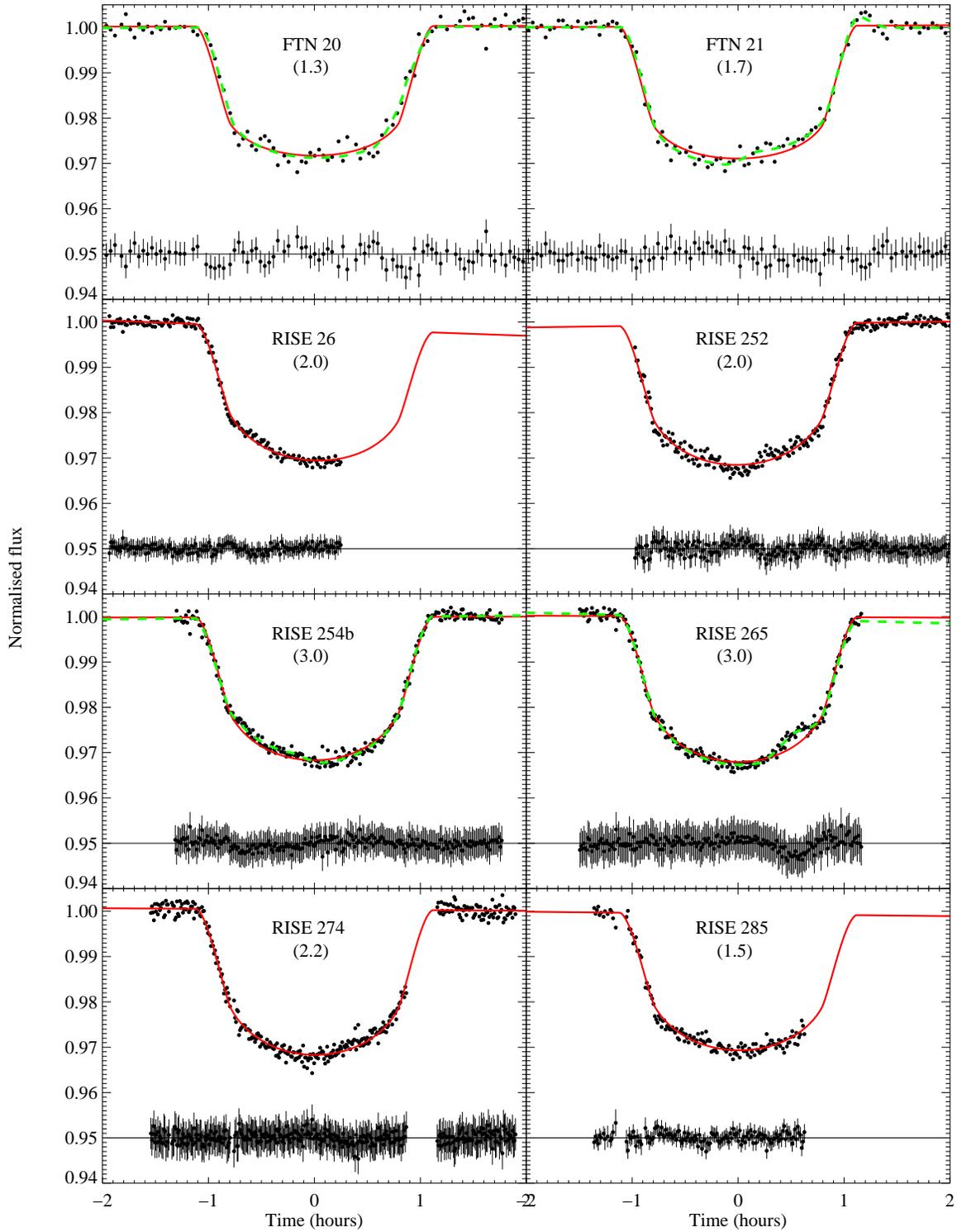}
  \caption{Phase-folded new light curves for WASP-10. From top to
    bottom and left to right in chronological order; FTN 2008 September 15 and 19, RISE 2008 October 04, RISE 2010 September 03 and 09, RISE 2010 October 13, RISE 2010 November 10, RISE 2010 December 14. The light curve name including the epoch number is printed in each plot. For each light curve, we superimpose the best-fit transit model 2 (solid red line), show the residuals from this model at the bottom of each the plot, and give the estimated red noise parametrised by $\beta$ inside parenthesis.
 For some light curves we also show a transit model that includes a spot feature (dashed green line). The light curves and models have been shifted to the T0 determined with model 2. The RISE data are binned into 45 second periods except for the 2010 November 10 which had a longer exposure time. The individual light curves  plotted here are available in electronic form at CDS.}
  \label{photolc}
\end{figure*}

The final new high precision transit light curves for WASP-10b are shown
in Figure~\ref{photolc} along with the best-fit model described in
Section~\ref{model}. We overplot the model residuals and the estimated
uncertainties, which are discussed in Section~\ref{errors}.

\begin{table*}
  \centering 
  \caption{WASP-10b new observations log.}
  \label{obser}
  \begin{tabular}{llcccccl}
    \hline
    \hline
Epoch &  Date & telescope & exp time & number exposures &  aperture radius  \\
    &   & sec  & & arcsec  \\
    \hline
20  &2008-09-15 & FTN  & 125 & 115 & 6.1 \\
21  &2008-09-19 & FTN  & 125 & 109 & 7.0\\
26  &2008-10-04 & RISE &  5  & 1580 & 8.6 \\
252 &2010-09-03 & RISE &  9  & 1195 & 8.1 \\
254b&2010-09-09 & RISE &  9  & 1233 & 8.6  \\
265 &2010-10-13 & RISE &  9  & 1063 & 10.3\\
274 &2010-11-10 & RISE & 34  & 339  & 11.4 \\
285 &2010-12-14 & RISE &  9  & 755  & 9.2 \\
    \hline
    \hline
  \end{tabular}
\end{table*}

Previously published WASP-10b light curves were included in our analysis which were kindly made available by the authors.
Details of these observations are given in Table~\ref{obser2}.  For completeness we also show the new best-fit model overploted on the previously published transit light curves in Figure~\ref{photolcA}, whose references for these data are given in Table~\ref{obser2}.

Hereafter, we refer to each light curve by their epoch number relative to the ephemeris presented in Section~\ref{ephe}. The epoch number is also given in the observation logs and light curve plots.

\begin{figure*}
  \centering
 \includegraphics[width=1.9\columnwidth]{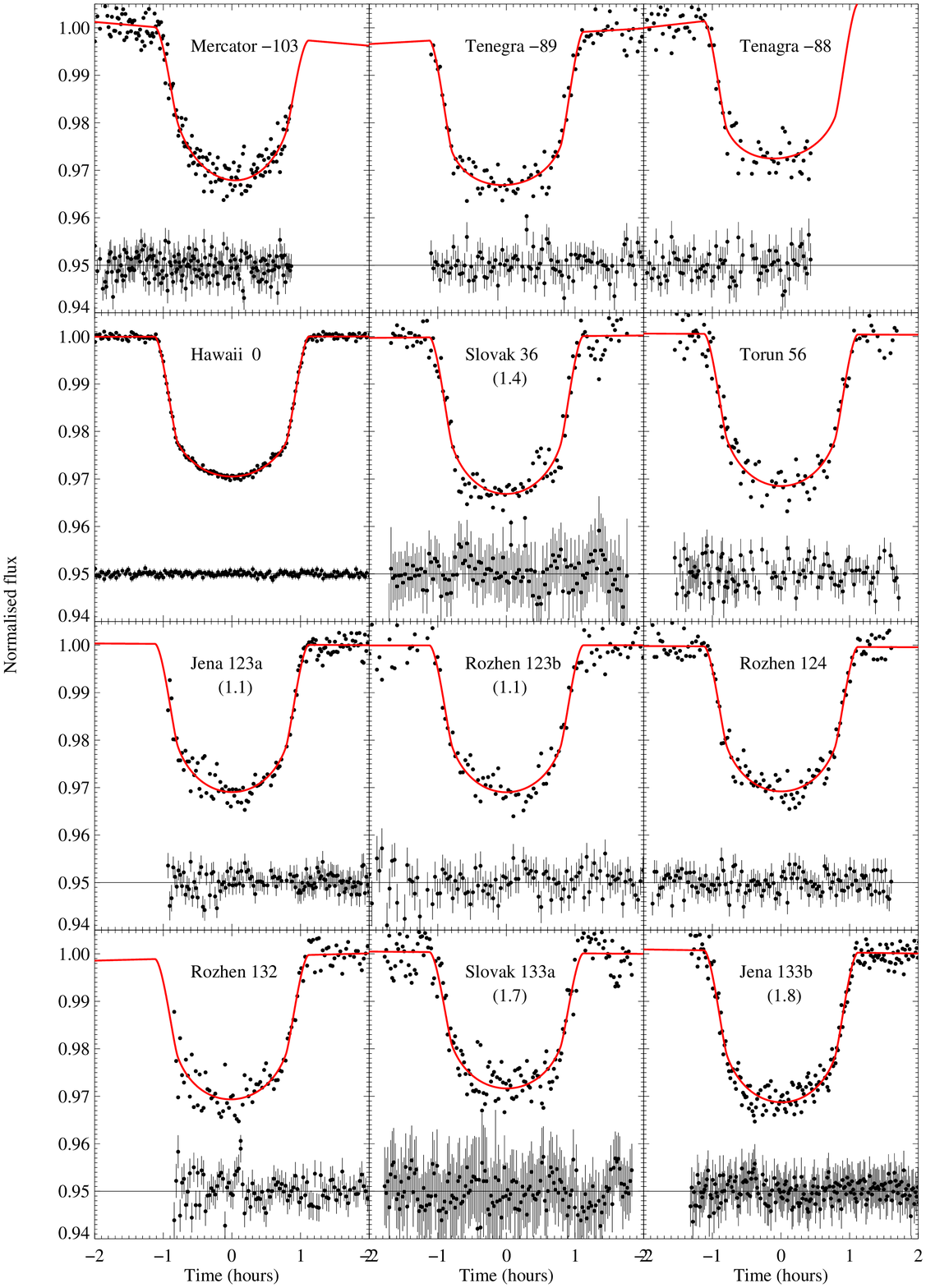}
  \caption{Phase-folded light curves for WASP-10 from previously published light curves which are described in Table~\ref{obser2}.
    Other comments are the same as Figure~\ref{photolc} but red noise values equal to 1 were omitted.}
  \label{photolcA}
\end{figure*}

\begin{figure*}
  \centering
 \includegraphics[width=1.9\columnwidth]{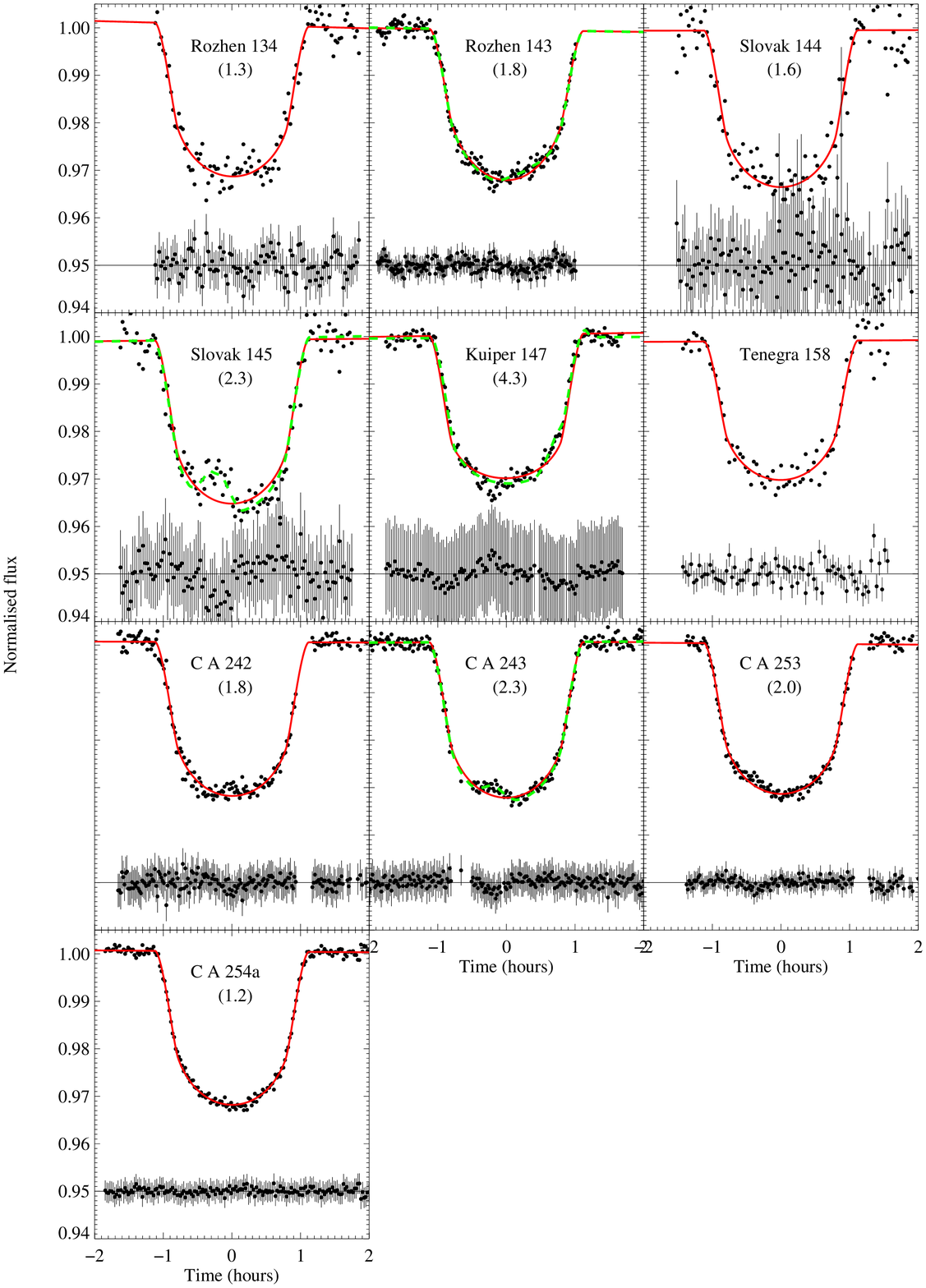}
  \contcaption{}
\end{figure*}

\begin{table*}
  \centering 
  \caption{WASP-10b previous observations log. The references are A = \citet{Christian2009}, B = \citet{Johnson2009}, C = \citet{Dittmann2010}, D = \citet{Krejcova2010}, E = \citet{Maciejewski2011}, and F = \citet{Maciejewski2011b}.}
  \label{obser2}
  \begin{tabular}{lclcccc}
    \hline
    \hline
 Epoch & Date & telescope & exp time & num exp & filter &  ref  \\
   & &   & sec  & & arcsec  \\
    \hline
-103 & 2007-09-01 & Mercator    & 30 & 170  & V & A\\
 -89 & 2007-10-15 & Tenagra     &125 & 125  & I & A\\
 -88 & 2007-10-18 & Tenagra     &125 &  91  & I & A\\
   0 & 2008-07-16 & 2.2m Hawaii &50  & 194  & sloan z'  & B \\ 
  36 & 2008-11-04 & Slovak      &    & 120  & R & D\\
  56 & 2009-01-05 & 0.6m Torun	& 50 & 254  & R & E\\
 123a& 2009-07-31 & 0.6m Jena 	& 70 & 144  & R & E\\	
 123b& 2009-07-31 & 0.6m Rozhen &120 & 116  & R & E\\
 124 & 2009-08-03 & 0.6m Rozhen & 90 & 122  & R & E\\
 132 & 2009-08-28 & 0.6m Rozhen & 90 &  99  & R & E\\
 133a& 2009-08-31 & Slovak      &    & 165  & R & D\\
 133b& 2009-08-31 & 0.6m Jena   & 55 & 203  & R & E\\
 134 & 2009-09-03 & 0.6m Rozhen & 90 & 105  & R & E\\
 143 & 2009-10-01 & 2.0m Rozhen & 15 & 548  & V & E\\
 144 & 2009-10-04 & Slovak      &    & 118  & R & D\\
 145 & 2009-10-07 & Slovak      &    & 114  & R & D\\
 147 & 2009-10-14 & Kuiper      &10  & 579  & I & C \\
 158 & 2009-11-17 & Tenagra 	& 76 &  76  & R & E\\
 242 & 2010-08-03 & 2.2m Calar Alto & 40 &165 &R &F \\
 243 & 2010-08-06 & 2.2m Calar Alto & 40 &213 &R &F \\
 253 & 2010-09-06 & 2.2m Calar Alto & 50 &164 &R &F \\
 254a& 2010-09-09 & 2.2m Calar Alto & 60 &184 &R &F \\
    \hline
    \hline
  \end{tabular}
\end{table*}

\section{Data analysis}

\subsection{Re-analyse of the radial velocity observations}

The discovery paper for WASP-10b \citep{Christian2009} favoured an eccentric orbit for the planet $e=0.059^{+0.014}_{-0.004}$ and $\omega=2.917^{+0.222}_{-0.245}$). However, \citet{Maciejewski2011} argued that eccentricity might have been overestimated due to stellar variability and favoured a circular orbit. Later, \citet{Husnoo2012} showed that the eccentricity detection was dependent on the first two radial velocity (RV) SOPHIE observations that could have a different zero point. Using a method of Bayesian model selection they find the eccentricity detection is not significant and obtained $e=0.052 \pm 0.031$. Furthermore,  the eccentricity is very correlated with the offset between the SOPHIE and FIES data. For all these reasons we assume a circular orbit. A better constraint on the eccentricity requires further RV observations.
 To estimate the stellar reflex velocity for a circular orbit we re-analysed the RV observations presented in \citet{Christian2009} using the new ephemeris given in Section~\ref{ephe}.

Using an MCMC routine we fitted 5 parameters of a keplerian model for the host star reflex motion: orbital period, central transit time (T0), RV semi-amplitude (K), centre-of-mass velocity ($\gamma$), and an offset between SOPHIE and FIES data.  We imposed a Gaussian prior to the orbital period and T0 using the new ephemeris given in Section~\ref{ephe}  and obtained  $K= 506.4 \pm 7.4\,ms^{-1}$, $\gamma=-11427.9 \pm 5.3\,ms^{-1} $, and offset$ = 112.0 \pm 1.2 \,ms^{-1}$. This solution has median rms of $43\,$m/s which compares with the median RV uncertainty of $20$ m/s. Hence, the radial velocities have extra jitter which could be due to stellar activity.

\begin{figure}
  \centering
  \includegraphics[width=\columnwidth]{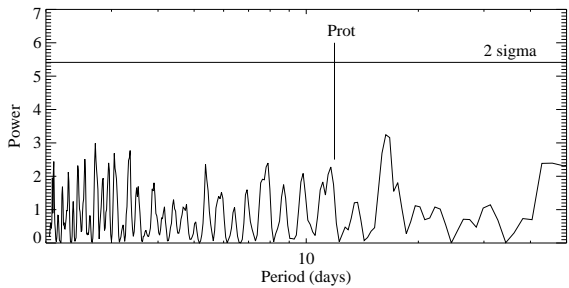}
  \caption{Lomb-Scargle periodogram of the RV residuals after subtracting a circular orbit for WASP-10b. The vertical line shows the rotation period of 11.91 and the horizontal line shows the false alarm probability at $2\, \sigma$ confidence limit. }
  \label{perio}
\end{figure}

\citet{Maciejewski2011} argued that the stellar activity had a sinusoidal effect (semi-amplitude of $65\,$m/s) in the measured RVs
 which should be accounted for when deriving the stellar reflex velocity due to WASP-10b. Following the procedure of \citet{Maciejewski2011},
  we computed the Lomb-Scargle periodogram of the RV residuals after removing the circular model given above. This is shown in Figure~\ref{perio} as well as the stellar rotation period of $11.91\pm 0.05\,$ d determined by \citet{Smith2009} and the false alarm probability at $2\,\sigma$ (95.4\%) confidence limit. We find no significant peak at the stellar rotation period, 
  and hence we cannot confirm that the radial velocities are modulated by spot variability. 
Moreover, other host stars that show 1-2\% photometric variability have smaller activity induced jitter in the measured RVs than the claimed $65\,$m/s for WASP-10. For example, HD 189733b has a RV jitter semi-amplitude of $15\,$m/s \citep{Boisse2009} and CoRoT-7b has a RV jitter semi-amplitude smaller than $20\,$m/s \citep{Queloz2009}.
Therefore, we conclude that the extra jitter observed in the RVs of WASP-10 does not appear to be periodic. Including one jitter parameter for each data set in the radial velocity fit we find that the FIES data jitter is less than $ 13\,ms^{-1}$ while the SOPHIE data jitter is $41 \pm 17\,ms^{-1}$. The latter is much higher than the expected jitter for SOPHIE supporting the result of \citet{Husnoo2012} about the possible offset of the two first SOPHIE points relative to the subsequent observations. Thus, we experimented discarding the first two SOPHIE points from the fit which resulted in a much lower estimated jitter for the SOPHIE data  $=13 \pm 18\,ms^{-1}$ compatible with the expected values for this instrument. This was adopted as our final solution for the RVs with the remaining parameters being: FIES jitter $  8 \pm 13 \,ms^{-1}$,  $K= 540 \pm 11\,ms^{-1}$, $\gamma=-11461  \pm 13\,ms^{-1} $, and offset$ = 84 \pm 18 \,ms^{-1}$. Our final solution for the stellar reflex velocity is in agreement with the value presented by \citet{Christian2009}.  The expected RV signal of the possible companion of $14\,$m/s is below the median RV uncertainty, as also pointed out by \citet{Maciejewski2011}.

% done in rvs/perio.pro

\label{rvs}

\subsection{Photometric errors}
\label{errors}
An accurate estimate of the photometric errors is important to obtain
reliable system parameters.  We begin by scaling the errors of each light curve so that
the $\chi_{\rm red}^2$  of the best fitting model is 1.0. This is especially important in our case because we are combining data from different instruments which were reduced with different pipelines.

Exoplanet transit observations are often affected by correlated red noise which must be taken into account in error analyses.
To estimate the time-correlated noise factor $\beta$ we followed the procedure of \citet{winn2008}. This consists of comparing the standard
 deviation of the residuals of the best fit solution for different time bins. For each light curve, we rescaled the errors 
 by the estimated $\beta$ which is given in Table~\ref{beta}.
 This procedure does not distinguish between systematic noise and real stellar variability.
 All departures from the transit shape are considered as noise. For example, when we fit simultaneously all
  light curves (model 2),  variations of the transit shape will also be considered as noise. Therefore,  the values of $\beta$ for the simultaneous fit are higher than for the individual fits.
 If the shape variations are real this procedure will overestimate the errors. However, if they are
  indeed caused by systematic noise it will allow a more accurate error estimation.

\begin{table}
  \centering 
  \caption{Estimated red noise values parametrised as $\beta$ calculated according to \citet{winn2008}}
  \label{beta}
  \begin{tabular}{lcc}
    \hline
    \hline
 Epoch & $beta$ ind & $beta$ simu\\
      \hline
 -103 &         1.0 &         1.0 \\  
    -89 &         1.0 &         1.0 \\  
    -88 &         1.0 &         1.0 \\  
      0 &         1.0 &         1.0 \\  
     20 &         1.0 &         1.3 \\  
     21 &         1.5 &         1.7 \\  
     26 &         1.5 &         2.0 \\  
     36 &         1.4 &         1.4 \\  
     56 &         1.0 &         1.0 \\  
    123a&         1.1 &         1.1 \\
    123b&         1.1 &         1.1 \\  
    124 &         1.1 &         1.0 \\  
    132 &         1.0 &         1.0 \\  
    133a&         1.4 &         1.7 \\  
    133b&         1.7 &         1.8 \\  
    134 &         1.3 &         1.3 \\  
    143 &         1.5 &         1.8 \\  
    144 &         1.4 &         1.6 \\  
    145 &         1.9 &         2.3 \\  
    147 &         2.4 &         4.3 \\  
    158 &         1.0 &         1.0 \\  
    242 &         1.5 &         1.8 \\  
    243 &         1.8 &         2.3 \\  
    252 &         2.0 &         2.0 \\  
    253 &         1.7 &         2.0 \\  
    254a&         1.0 &         1.2 \\ 
    254b&         3.0 &         3.0 \\  
    265 &         2.7 &         3.0 \\  
    274 &         1.8 &         2.2 \\  
    285 &         1.3 &         1.5 \\  
  \hline
    \hline
  \end{tabular}
\end{table}

\subsection{Transit model and fitting procedure}
\label{model}

To determine the planetary and orbital parameters, we fitted the light curves of WASP-10b
with a transit model coupled with an Markov Chain Monte Carlo (MCMC)
 routine, a procedure similar to \citet{Barros2011b} for WASP-21b.
In our previous transit fitting analyses we fit all the light curves simultaneously. However, here we also present the individual light curve fits, mostly for comparison with previous results and call the latter model 1 and the former model 2.
 We used the
\citet{Mandel2002} transit model parametrised by the normalised
separation of the planet, $a/R_*$, ratio of planet radius to star
radius, $ R_p/R_* $, orbital inclination, $i$, and the transit epoch,
$T_0$, of each light curve. A circular orbit was adopted as discussed above. For each light curve, we also include two extra
parameters ($F_{out}$  and $T_{grad}$) to account for a linear normalisation.
 We used the quadratic limb darkening coefficients (LDCs)  deduced from the tables of \citet{Howarth2011a} 
 for $T_{eff} = 4675 \,$K, \logg $ = 4.4$ and [M/H] $ = 0.03$ \citep{Christian2009}. Most of the light curves are of 
 insufficient quality to allow for the fitting of the LDCs. To be consistent in the treatment of all the light
  curves we fixed the limb darkening coefficients to the theoretical values. Although this may lead to a slight underestimation of the errors, it was preferred to avoid spurious correlations. The values of the LDCs used for each filter are given in Table~\ref{ldcoef}.
 
\begin{table}
  \centering 
  \caption{Limb darkening coefficients used for each filter.}
  \label{ldcoef}
  \begin{tabular}{lcccl}
    \hline
    \hline
  filter & linear coef $a$ &  quadratic coef $b$\\
 sloan z'&  0.391 & 0.210\\ 
  I      &  0.421 & 0.192\\
  R      &  0.602 & 0.133\\ 
  rise   &  0.612 & 0.146\\
  V      &  0.763 & 0.036\\
    \hline
    \hline
  \end{tabular}
\end{table}

The MCMC chains were constructed following the Metropolis-Hastings algorithm. At each step of the chain each parameter is perturbed by a jump function, which is a random number sampled from a Gaussian distribution with mean zero and standard deviation equal to the respective parameter uncertainty. The new parameter set is accepted with probability: $P=min \left( 1, exp\left(\frac{-\Delta\chi^2}{2}\right) \right)$where $\Delta\chi^2$ is the difference in the $\chi^2$ of subsequent parameters sets. The jump functions are scaled by a common factor in order to ensure that the percentage of accepted steps lies between $20\%$ and $30\%$. Further details about the fitting procedure can be found in  \citet{Barros2011b}. 
For both models the initial 20\% of each MCMC chain that
corresponded to the burn-in phase were discarded and the remaining
parts merged into a master chain.  The best fit parameter was estimated
as the median of its probability distribution and the  $1\,\sigma$ limits
as the value at which the integral of the distribution equals 0.341
from both sides of the median.  To test convergence, the \citet{Gelman92} statistic was computed for each fitted parameter and was found to be less than $0.9\%$ from unity for all the parameters in both models.

\subsubsection{Model 1}
The existence of an additional planet or moon affects the transit timings of the transiting planet, but can also affect the orbital parameters of the system.
Hence, in previous analyses of the transit times of WASP-10, the authors opted for fitting each light curve individually \citep{Dittmann2010, Maciejewski2011, Maciejewski2011b}.
 However, in some cases they were forced to fix the inclination of the system due to the poor quality of individual light curves.
Here, we follow the same procedure primarily for comparison with previous methods. The results will also test possible variations of the light curve shape.
Thus, we fitted each light curve individually using the MCMC method explained above with $500\, 000$ points per chain.  Five chains were combined to obtain the final result.
Instead of fixing the inclination we choose to impose a prior of the form:
  \begin{equation}
    \frac{(i-i_{0})^2}{\sigma^2_{i}},
  \end{equation}
  where we assume $i_{0}=88.66, \sigma_{i}=0.12$, taken from the simultaneous fit (model 2).
  This allows to account for the uncertainty of the inclination in the fit. We refer to this as model 1.

\subsubsection{Model 2}

For model 2 we assume that a possible additional planet has no significant effect on the transit shape during the time span of the observations. 
Therefore, we performed a simultaneous MCMC fit to all 30 light curves, fitting globally $a/R_*$, $ R_p/R_* $, and  $i$
and individually $T0$,  $F_{out}$ and $T_{grad}$. We remind the reader that deviations from a constant shape are considered to be only due to systematic noise or intrinsic stellar variability and hence are treated as noise. Specifically, deviations of limb darkening parameters from the tabulated values are also considered as noise.
 However, since the brightness of WASP-10 has been shown to vary significantly due to star spots \citep{Smith2009,Maciejewski2011} a factor $df$ was included to account for out-of-transit flux variability between each light curve.
  This factor has been set arbitrarily to zero for our best quality light curve (Epoch 0, \citet{Johnson2009}) which is also the one taken in the reddest filter, and hence that should be less affected by stellar variability. Therefore, a total of $30\times 4+3=123$ parameters were fitted simultaneously.
We computed seven MCMC chains each of  $1\, 500\, 000$ points and different initial parameters.

\section{Results}

\subsection{Model 1}

The individual light curve fitted parameters are given in Table~\ref{resm1}. Transit times were converted to Barycentric Dynamical Time (TDB) using the IDL codes kindly made available by \citet{Eastman2010}.
  The weighted mean of the fitted parameters is: $a/R_* = 11.910 \pm  0.024$, $ R_p/R_* =  0.15869  \pm 0.00019$, and  $i = 88.67$ degrees and the dispersion of inclination is equal to $0.13\,$ degrees as expected from the prior.

Comparing the values of $a/R_*$ for all light curves we conclude that individual deviations from the mean value of the normalised
separation of the planet are smaller than $2\sigma$ except for 2 transits. These are epoch 20 which was observed in  2008-09-15 with FTN and epoch 147 \citep{Dittmann2010}. Both have $a/R_*$ higher than the mean  $a/R_*$ by more than $3\sigma$.
Both of these transits show signs of correlated noise either due to systematics or stellar variability and will be discussed in section 5.2.

As expected, due to spot induced brightness variability of the host star, the parameter $ R_p/R_*$ shows a wide spread.
Two of the light curves (243, 254a) show a significantly higher ($3\,\sigma$) value of $R_p/R_*$, while three light curves (0,133,158) show a significantly lower value. 
Therefore, we confirm transit depth variations reported by other authors and the need to include the extra $df$ parameter in the global fit. 
 For completeness we also tested a model with free inclination and concluded that the inclination is constant within the uncertainties.
 Hence, we conclude that there are no significant shape variations apart from the transit depth,  except for transits epoch 20 and 147.

Besides testing variations of transit shape, model 1 is also useful for direct comparison with previous transit times.  This allows to better isolate the causes for the differences in the estimated transit times. 
%In Figure~\ref{peri_ind} we show the difference between our derived transit times and previously published results. 
The first three transits (-103,-89,-88) were previously combined together with the WASP data in a single ephemeris, and hence, they cannot be directly compared.
 Regarding the other previously published transit times, our results are within $1\sigma$ of previous results except for epochs 36 and 147 for which the difference is respectively  $-2.0 \pm 0.94$ and  $0.67 \pm 0.47$ minutes.
The disagreement can be explained by failure to account for the baseline function in the previous analysis of these transits. In our model a linear trend is fitted simultaneously with the transit model.  When we do not fit a trend we recover the previous published results. T0 correlates with the baseline function because the latter affects the symmetry of the light curve. Therefore, including the baseline function is crucial to obtain reliable T0s

%% We suspect this is due to trends present in the data what were not accounted for in the previous analyses (see caption of Figure 2  of \citet{Maciejewski2011}).

 In general, we obtain slightly larger errors ($\sim 1.6\times$) than previous studies. For each case this is due to one or more of the following differences from the previous results: inclusion of a baseline function in the fit, inclusion of the red noise factor, and/or the treatment of the inclination. 
 We argue that fixing or artificially constraining the inclination will cause underestimation of the transit time errors. %For example, for epoch 0 our errors are artificially smaller compared to \citet{Johnson2009} where the inclination remained free.

 Despite the difference from previous results being small, the new transit times are in much better agreement with a linear ephemeris. The  $\chi^2$ of a linear fit to the transit times is 91.95 for 28 degrees of freedom, i.e. $\chi_{\rm red}^2=3.28$ which compares with $13.8$ found by \citet{Maciejewski2011}. However, the p-value is still low $9.777 \times 10^{-9}$  and hence the null hypothesis can be discarded at 5.7 $\sigma$.  Four of the new transit times deviate more than $3\, \sigma$ from a linear ephemeris. These are epochs 20, 21, 143, and 147. Both epoch 21 and epoch 147 show trends in the light curve that can be affecting the transit times.

 To test the model for the TTVs proposed by \citet{Maciejewski2011} we fitted the sinusoidal model given by equation 1 of \citet{Maciejewski2011} for frequencies: $f_{1}^{ttv} = 0.183\, cycl\, P^{-1}$ and $f_{2}^{ttv} =  0.175\, cycl\, P^{-1}$, where P is the orbital period of WASP-10b. 
The preferred solution presented by \citet{Maciejewski2011} for frequency $f_{1}^{ttv}$ had a p-value equal to  $3.4 \times 10^{-4}$ which suggests that the model could be rejected at the $3.6\sigma$ level. However, we remind the reader that this sinusoidal model is only used to estimate the most probable TTV frequency to be used later in the 3-body simulations. Therefore, the p-value test cannot be directly applied. Nevertheless, we found that the sinusoidal model fit to the new transit times is worse, as it results in a much lower p-value $4.1 \times 10^{-7}$ for both frequencies $f_{1}^{ttv}$ and $f_{2}^{ttv}$. Most importantly, for the new results we obtain 3.5 times smaller TTV semi-amplitudes:  $A_{1}^{ttv}= 0.48 \pm 0.18$ and $A_{2}^{ttv}= 0.53 \pm 0.14$.

\begin{figure}
  \centering
  \includegraphics[width=\columnwidth]{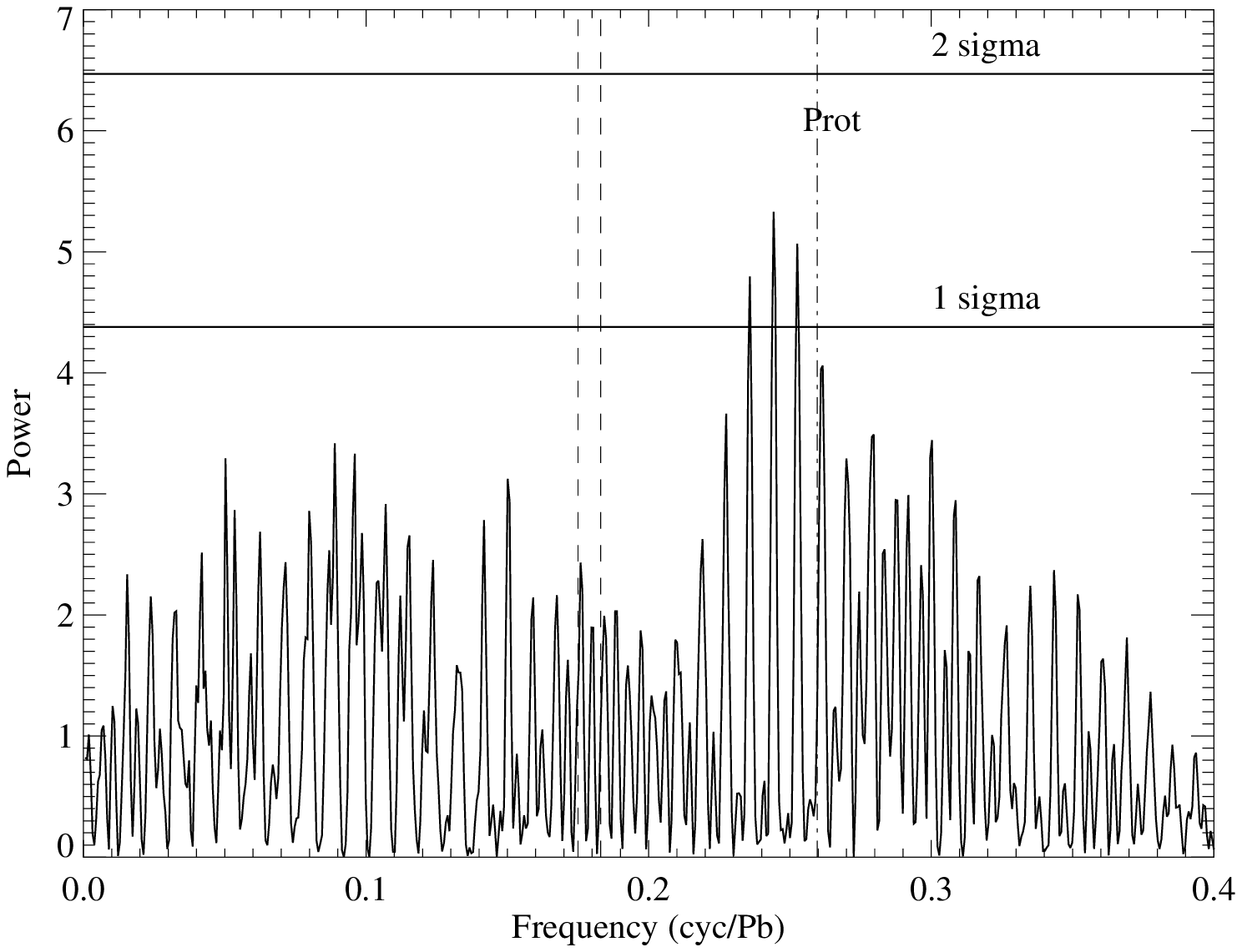}
  \caption{Lomb-Scargle periodogram of residual transit times from model 1 after subtracting a linear ephemeris. The vertical dash-dotted line shows the rotation  period of 11.91. The dashed lines show the TTV frequencies proposed by \citet{Maciejewski2011} and the horizontal line shows the false alarm probability at 1 and $2\,\sigma$ confidence limit. }
  \label{peri_ind}
\end{figure}

In Figure~\ref{peri_ind} we show the Lomb-Scargle periodogram of the residual transit times of model 1 after subtracting a linear ephemeris. 
Notice that the TTV frequencies proposed by \citet{Maciejewski2011} have disappeared which agrees with the much lower amplitudes found for the sinusoidal model fit.

\begin{table*}
  \centering 
  \caption{WASP-10 MCMC fitted parameters for model 1 (individual fit). Times are in TDB.}
  \label{resm1}
   \begin{tabular}{cccccc}
    \hline
    \hline
Number  &  $a/R_*$ & $ R_p/R_* $ &$inc $ (degrees) & $T_{1-4}$  (hours) &T0 (days)\\
 -103 &      12.02 $\pm$        0.21  &     0.1509 $\pm$      0.0034  &      88.67 $\pm$        0.12  &      2.198 $\pm$       0.043 &      4345.48630 $\pm$          0.00061\\
   -89 &      12.02 $\pm$        0.17  &     0.1620 $\pm$      0.0030  &      88.65 $\pm$        0.12  &      2.219 $\pm$       0.033 &      4388.78602 $\pm$          0.00042\\
   -88 &      11.13 $\pm$        1.06  &     0.1626 $\pm$      0.0046  &      88.67 $\pm$        0.12  &      2.411 $\pm$       0.245 &      4391.88193 $\pm$          0.00416\\
     0 &      11.91 $\pm$        0.06  &     0.1575 $\pm$      0.0003  &      88.70 $\pm$        0.10  &      2.236 $\pm$       0.005 &      4664.03808 $\pm$          0.00005\\
    20 &      12.35 $\pm$        0.13  &     0.1557 $\pm$      0.0011  &      88.64 $\pm$        0.12  &      2.143 $\pm$       0.017 &      4725.89175 $\pm$          0.00026\\
    21 &      12.01 $\pm$        0.14  &     0.1570 $\pm$      0.0014  &      88.67 $\pm$        0.12  &      2.212 $\pm$       0.023 &      4728.98414 $\pm$          0.00038\\
    26 &      11.15 $\pm$        0.41  &     0.1545 $\pm$      0.0021  &      88.72 $\pm$        0.11  &      2.392 $\pm$       0.097 &      4744.45206 $\pm$          0.00160\\
    36 &      11.96 $\pm$        0.18  &     0.1641 $\pm$      0.0023  &      88.66 $\pm$        0.12  &      2.235 $\pm$       0.033 &      4775.37693 $\pm$          0.00051\\
    56 &      11.93 $\pm$        0.15  &     0.1602 $\pm$      0.0017  &      88.65 $\pm$        0.12  &      2.233 $\pm$       0.027 &      4837.23105 $\pm$          0.00042\\
   123 &      11.93 $\pm$        0.16  &     0.1572 $\pm$      0.0019  &      88.66 $\pm$        0.12  &      2.227 $\pm$       0.029 &      5044.44435 $\pm$          0.00044\\
   123 &      11.90 $\pm$        0.16  &     0.1580 $\pm$      0.0026  &      88.68 $\pm$        0.12  &      2.238 $\pm$       0.031 &      5044.44407 $\pm$          0.00041\\
   124 &      11.86 $\pm$        0.12  &     0.1559 $\pm$      0.0013  &      88.67 $\pm$        0.12  &      2.239 $\pm$       0.020 &      5047.53701 $\pm$          0.00033\\
   132 &      11.67 $\pm$        0.51  &     0.1582 $\pm$      0.0057  &      88.66 $\pm$        0.12  &      2.281 $\pm$       0.116 &      5072.27832 $\pm$          0.00182\\
   133 &      11.85 $\pm$        0.15  &     0.1598 $\pm$      0.0019  &      88.67 $\pm$        0.12  &      2.249 $\pm$       0.027 &      5075.37131 $\pm$          0.00041\\
   133 &      12.11 $\pm$        0.21  &     0.1511 $\pm$      0.0024  &      88.67 $\pm$        0.12  &      2.181 $\pm$       0.038 &      5075.37145 $\pm$          0.00063\\
   134 &      11.75 $\pm$        0.23  &     0.1623 $\pm$      0.0043  &      88.66 $\pm$        0.12  &      2.274 $\pm$       0.051 &      5078.46453 $\pm$          0.00080\\
   143 &      11.76 $\pm$        0.14  &     0.1576 $\pm$      0.0024  &      88.68 $\pm$        0.11  &      2.265 $\pm$       0.028 &      5106.29690 $\pm$          0.00038\\
   144 &      11.62 $\pm$        0.24  &     0.1629 $\pm$      0.0035  &      88.67 $\pm$        0.12  &      2.304 $\pm$       0.049 &      5109.39109 $\pm$          0.00080\\
   145 &      11.50 $\pm$        0.26  &     0.1649 $\pm$      0.0034  &      88.67 $\pm$        0.12  &      2.334 $\pm$       0.055 &      5112.48544 $\pm$          0.00088\\
   147 &      12.41 $\pm$        0.13  &     0.1601 $\pm$      0.0013  &      88.65 $\pm$        0.12  &      2.142 $\pm$       0.019 &      5118.66785 $\pm$          0.00030\\
   158 &      12.00 $\pm$        0.15  &     0.1529 $\pm$      0.0018  &      88.66 $\pm$        0.12  &      2.206 $\pm$       0.026 &      5152.68920 $\pm$          0.00041\\
   242 &      11.83 $\pm$        0.11  &     0.1608 $\pm$      0.0010  &      88.67 $\pm$        0.12  &      2.256 $\pm$       0.016 &      5412.47863 $\pm$          0.00024\\
   243 &      11.81 $\pm$        0.09  &     0.1616 $\pm$      0.0009  &      88.68 $\pm$        0.12  &      2.264 $\pm$       0.013 &      5415.57161 $\pm$          0.00020\\
   252 &      12.03 $\pm$        0.14  &     0.1550 $\pm$      0.0024  &      88.68 $\pm$        0.12  &      2.205 $\pm$       0.026 &      5443.40659 $\pm$          0.00035\\
   253 &      11.80 $\pm$        0.09  &     0.1592 $\pm$      0.0009  &      88.68 $\pm$        0.11  &      2.260 $\pm$       0.013 &      5446.49881 $\pm$          0.00018\\
   254 &      11.86 $\pm$        0.08  &     0.1610 $\pm$      0.0005  &      88.62 $\pm$        0.11  &      2.245 $\pm$       0.007 &      5449.59138 $\pm$          0.00007\\
   254 &      11.86 $\pm$        0.12  &     0.1581 $\pm$      0.0014  &      88.66 $\pm$        0.12  &      2.245 $\pm$       0.020 &      5449.59120 $\pm$          0.00031\\
   265 &      11.71 $\pm$        0.17  &     0.1614 $\pm$      0.0026  &      88.66 $\pm$        0.12  &      2.281 $\pm$       0.035 &      5483.61165 $\pm$          0.00054\\
   274 &      11.92 $\pm$        0.10  &     0.1596 $\pm$      0.0008  &      88.67 $\pm$        0.12  &      2.236 $\pm$       0.014 &      5511.44560 $\pm$          0.00021\\
   285 &      11.48 $\pm$        0.53  &     0.1573 $\pm$      0.0038  &      88.69 $\pm$        0.12  &      2.323 $\pm$       0.122 &      5545.46688 $\pm$          0.00205\\
  \hline
     \hline
  \end{tabular}
\end{table*}

 \label{resmodel1}

\subsection{Model 2}

The individual transit times and the out-of-transit flux variability derived from a simultaneous fit of the 30 light curves of WASP-10 are given in Table~\ref{resm2}.

\begin{table*}
  \centering 
  \caption{WASP-10 MCMC fitted parameters for model 2 (simultaneous fit). We also give the time residuals from a linear ephemeris in seconds.}
   \label{resm2}
  \begin{tabular}{lccc}
    \hline
    \hline
Number  &  df  & T0    & Time residuals \\ 
        &      &  days &  sec \\
  -103 &      0.047 $\pm$       0.065  & 4345.48665  $\pm$     0.00038 &      -28.4 $\pm$        32.6\\
   -89 &     -0.067 $\pm$       0.030  & 4388.78596  $\pm$     0.00041 &       66.7 $\pm$        35.7\\
   -88 &     -0.040 $\pm$       0.057  & 4391.87917  $\pm$     0.00054 &      108.0 $\pm$        46.4\\
     0 &      0.000 $\pm$       0.000  & 4664.03808  $\pm$     0.00005 &       -2.4 $\pm$         4.6\\
    20 &      0.030 $\pm$       0.032  & 4725.89179  $\pm$     0.00038 &      -77.9 $\pm$        32.4\\
    21 &      0.007 $\pm$       0.020  & 4728.98421  $\pm$     0.00043 &     -104.6 $\pm$        37.3\\
    26 &      0.065 $\pm$       0.024  & 4744.44923  $\pm$     0.00021 &       13.9 $\pm$        18.3\\
    36 &     -0.066 $\pm$       0.053  & 4775.37696  $\pm$     0.00054 &       51.0 $\pm$        46.7\\
    56 &     -0.029 $\pm$       0.021  & 4837.23102  $\pm$     0.00043 &        5.8 $\pm$        37.4\\
   123a&     -0.002 $\pm$       0.028  & 5044.44407  $\pm$     0.00031 &       20.8 $\pm$        26.9\\
   123b&      0.004 $\pm$       0.024  & 5044.44434  $\pm$     0.00046 &       44.1 $\pm$        39.3\\
   124 &      0.018 $\pm$       0.016  & 5047.53700  $\pm$     0.00030 &       38.0 $\pm$        26.3\\
   132 &      0.034 $\pm$       0.044  & 5072.27905  $\pm$     0.00052 &       56.7 $\pm$        45.0\\
   133a&      0.080 $\pm$       0.034  & 5075.37113  $\pm$     0.00073 &        0.9 $\pm$        62.7\\
   133b&     -0.022 $\pm$       0.025  & 5075.37133  $\pm$     0.00043 &       18.2 $\pm$        37.4\\
   134 &     -0.028 $\pm$       0.036  & 5078.46489  $\pm$     0.00062 &       90.3 $\pm$        53.5\\
   143 &      0.023 $\pm$       0.020  & 5106.29655  $\pm$     0.00032 &     -161.2 $\pm$        27.9\\
   144 &     -0.063 $\pm$       0.053  & 5109.39103  $\pm$     0.00095 &      -10.2 $\pm$        82.0\\
   145 &     -0.110 $\pm$       0.069  & 5112.48544  $\pm$     0.00110 &      135.1 $\pm$        95.4\\
   147 &     -0.026 $\pm$       0.029  & 5118.66795  $\pm$     0.00056 &     -119.4 $\pm$        48.8\\
   158 &      0.057 $\pm$       0.020  & 5152.68916  $\pm$     0.00044 &      -16.6 $\pm$        37.7\\
   242 &     -0.043 $\pm$       0.014  & 5412.47858  $\pm$     0.00028 &       -4.0 $\pm$        24.2\\
   243 &     -0.053 $\pm$       0.013  & 5415.57163  $\pm$     0.00026 &       23.4 $\pm$        22.4\\
   252 &      0.009 $\pm$       0.021  & 5443.40631  $\pm$     0.00027 &       33.5 $\pm$        22.9\\
   253 &     -0.018 $\pm$       0.012  & 5446.49884  $\pm$     0.00021 &       16.1 $\pm$        18.4\\
   254b &     -0.010 $\pm$       0.021  & 5449.59123  $\pm$     0.00031 &      -12.8 $\pm$        27.0\\
   254a &     -0.040 $\pm$       0.006  & 5449.59139  $\pm$     0.00009 &        0.7 $\pm$         8.1\\
   265 &     -0.025 $\pm$       0.026  & 5483.61125  $\pm$     0.00044 &      -13.7 $\pm$        38.3\\
   274 &     -0.025 $\pm$       0.011  & 5511.44562  $\pm$     0.00025 &      -29.9 $\pm$        21.8\\
   285 &      0.037 $\pm$       0.025  & 5545.46534  $\pm$     0.00029 &      -56.4 $\pm$        25.2\\
  \hline
     \hline
  \end{tabular}
\end{table*}

We conclude that three light curves 242, 243, 254a (\citealt{Maciejewski2011b}) have a significantly lower df (at $3\, \sigma$ ) than the chosen zero point. In our notation a negative df means the star is fainter which corresponds to a larger transit depth.  
Some light curves have a higher df, but this is non-significant. Therefore, we confirm that our chosen zero point is valid.  
 However, this zero point might not correspond to the real maximum brightness of WASP-10, and hence the planet to star ratio could still be overestimated.

\begin{figure}
  \centering
  \includegraphics[width=\columnwidth]{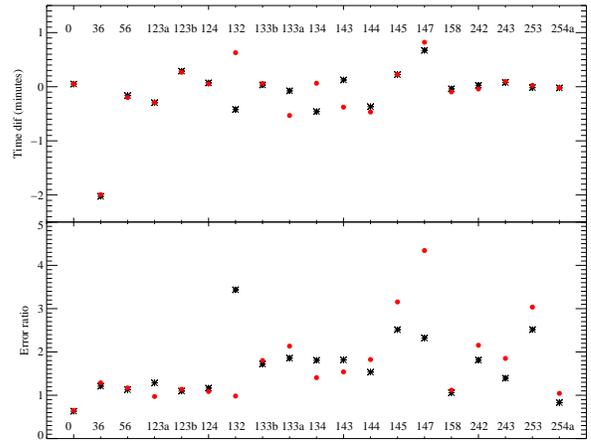}
  \caption{Comparison between our results and previously published ones. We show the time difference in the top panel and the error ratio in the bottom panel. Model 1 comparison is shown as black stars and model 2 as red circles.}
  \label{comptimes}
\end{figure}

 In Figure~\ref{comptimes} we plot the difference between the our transit times derived with model 1 (black stars) and model 2 (red dots) and previously published transit times. We also show the ratio of transit time errors. The highest difference in times is found for transits 36, 132 and 147. For epochs 36 and 147 this is due to the baseline function not been accounted for in previous fits, as mentioned before. For epoch 132 the difference is within the errors and it is due to epoch 132 being a partial transit and hence very sensitive to the details of the model fit. 
The transit times derived from model 1 and model 2 are consistent within $1\, \sigma$. The biggest differences in the results from the two models are in partial transits.
In model 2, the global shape of the transit defined by the complete transits constrains the shape of partial transits,  this allows a better constrain on the baseline function that correlates with the transit times.

 From the bottom panel of Figure~\ref{comptimes} it is also clear that our estimated errors are higher than previous ones. This is due to the inclusion of red noise in our analysis.
For model 2 the uncertainties of T0s of partial transits are smaller than for model 1 because the shape of the transit in model 2 is almost fixed by the other light curves. 
In contrast, some T0 uncertainties of complete transits are higher because differences from the common shape are considered red-noise which is accounted for as explained in Section~\ref{errors}.
This is the key difference between the models and implies that in this case model 2 will produce much more reliable results, as will be discussed later.

\begin{figure*}
  \centering
  \includegraphics[width=15cm]{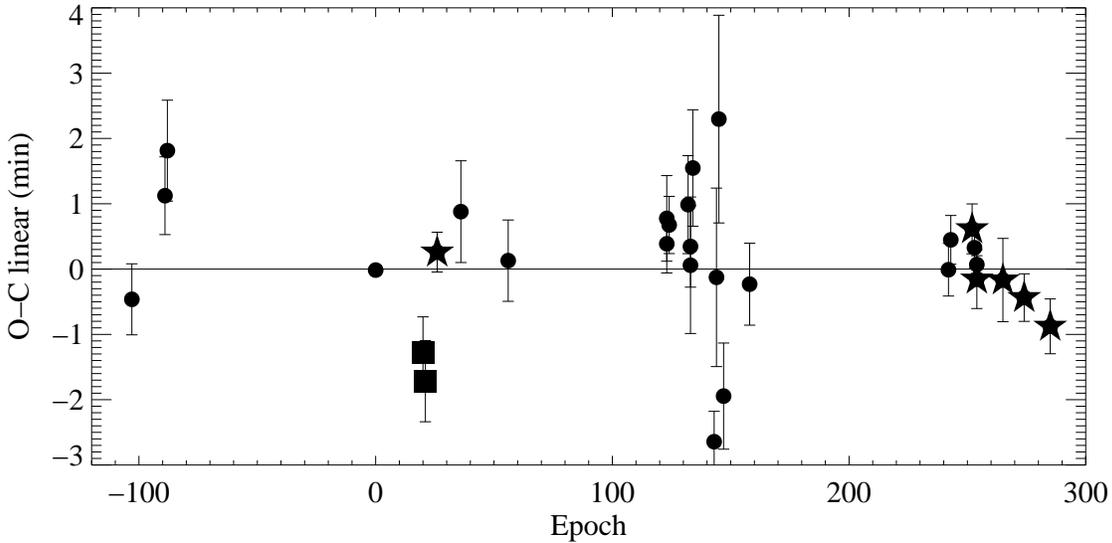}
  \caption{ Model 2 estimated transit residuals of WASP-10b after subtracting a linear ephemeris. The transit times were derived from a simultaneous fit of all the light curves. Previously published fitted light curves are shown as circles. The new RISE data are shown as stars while the new FTN data as squares.}
  \label{timeplot}
\end{figure*}

We compute a linear ephemeris using the new estimated transit times for model 2:
\begin{equation}
  T_{t} (TDB) = T(0) + EP,
\end{equation}
where $P=3.09272932 \pm  0.00000032 $ and $T_0 =2454664.038090 \pm
 0.000048$. Time residuals (or TTVs) from the linear ephemeris are given in Table~\ref{resm2} and shown in Figure~\ref{timeplot}. The RISE data are show as stars, the new FTN data as squares, and the new results based on previously published light curves as circles.

\label{ephe}

%the mean rms of the residuals is 45 seconds. 
  The  $\chi^2$  of the linear ephemeris is 86 for 28 degrees of freedom and the p-value is $5.7 \times 10^{-8}$ so we conclude that the linear fit is not a a good fit at $5.4 \sigma$ confidence level. However, only epoch 143 deviates significantly (at $5.8\, \sigma$) from the linear ephemeris. This transit is the only one obtained with the 2.0m Rozhen telescope and is a partial transit. If we discard epoch 143 the  $\chi^2$ decreases to 53 for 27 degrees of freedom and the p-value increases to 0.00176 and hence we can only reject the null hypotheses at the $3.1\sigma$ level.

 For these new transit times derived with model 2 we also tested the sinusoidal model proposed by \citet{Maciejewski2011} for the two proposed frequencies as done in the previous section for model 1 and found similar results. We find  $A_{1}^{ttv}= 0.52 \pm 0.17$ and $A_{2}^{ttv}= 0.45 \pm 0.14$ while both p-values are less then $1.6 \times 10^{-6}$ and hence the model could be rejected at the $4.8 \sigma$ level.  We reiterate that the expected TTV variations are not strictly periodic and the p-value can not be directly applied. However, clearly the new results do not support the TTV solution claimed by \citet{Maciejewski2011}. We analyse a perturber model for the our derived transit times in Section~\ref{masslimit}.

\begin{figure}
  \centering
  \includegraphics[width=\columnwidth]{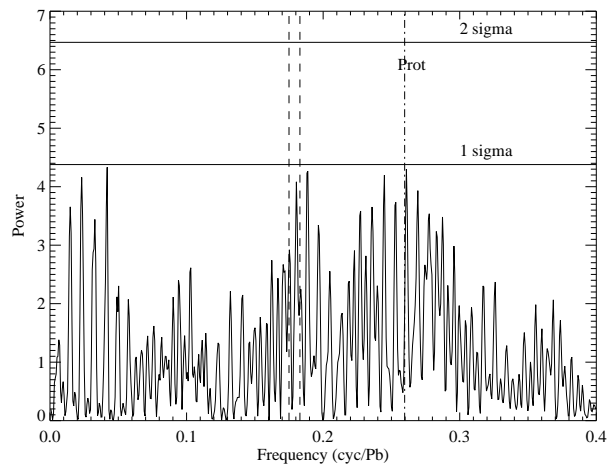}
  \caption{Lomb-Scargle periodogram of model 2 residual transit times after subtracting a linear ephemeris. The vertical dash-dotted line shows the rotation period of 11.91 the dashed lines show the TTV frequencies proposed by \citet{Maciejewski2011} and the horizontal lines show the false alarm probability at 1 and $2\, \sigma$ confidence limit. }
  \label{peri_simu}
\end{figure}

The periodogram of the time residuals relative to a linear ephemeris are given in Figure~\ref{peri_simu}. 
Comparing with the model 1 results, it is clear that the peaks around the star's rotation period have decreased, implying that model 2 results are less affected by stellar variability.

\subsection{System parameters}

Although the main objective of this study is not deriving updated parameters for WASP-10, for consistency with the simulations presented in Section~\ref{masslimit} we derive the system parameters from the geometric fitted parameters, given in the top of Table~\ref{MCMC}.
The same procedure as \citet{Barros2012} is followed.

We  estimate the stellar mass from models by interpolating
the Yonsei-Yale stellar evolution tracks by \citet{Demarque2004} using
the effective temperature $T_{eff}=4675 \pm 100\,$K and metallicity  ${[M/H]} =0.03 \pm 0.2\, $dex from \citet{Christian2009} 
and our new derived stellar density $ \rho_*  = 2.359^{+0.053}_{-0.047} \rho_{\odot} $.
We obtain a sightly larger stellar density than derived by \citet{Johnson2009} ( $ \rho_*  = 2.20 \pm 0.06 \rho_{\odot} $) which results in a slightly lower mass  $0.73 \pm 0.1 \Msun$, but consistent with previous results.

Combining the stellar mass, stellar reflex velocity estimated in Section~\ref{rvs}, and the geometric parameters derived with model 2, we derive the system parameters which are given on Table~\ref{MCMC}. 
The parameters were derived using the equations of \citet{Seager2003} and \citet{Kipping2010}. Specifically, the $T_{1-4}$ refers to the total transit duration and $T_{T1}$  is the duration of the sky-projected centre of the planet overlapping the stellar surface and  is defined by equation 15 of \citet{Kipping2010}.

\begin{table}
  \centering 
  \caption{WASP-10 system parameters derived from model 2. The first set of parameters are derived directly from the transit light curves.}
  \label{MCMC}
  \begin{tabular}{lccl}
    \hline
    \hline
    Parameter & Value   \\
    \hline
    Normalised separation $a/R_*$ & $  11.895 \pm 0.083$  \\
    Planet/star radius ratio $ R_p/R_* $ & $ 0.15758^{+0.00036}_{-0.00039}$\\
    Transit duration $T_{1-4}$ [hours] & $  2.2363^{0.0051}_{-0.0051}$ \\
    Transit duration $T_{T1}$ [hours] & $ 1.9110^{0.0020}_{-0.0021}$ \\
    Orbital inclination $ I $ [degrees] & $   88.66 \pm 0.12$ \\
    Impact parameter $ b $ [$R_*$]  & $   0.277^{+0.021}_{-0.025}$ \\
    Stellar density $ \rho_* $ [$\rho_{\odot}$] & $ 2.359^{+0.053}_{-0.047}  $ \\
    &    &      &  \\
    Orbital semimajor axis $ a $ [AU] & $  0.0375\pm 0.0017 $ \\
    Stellar mass $ \mathrm{M}_* $ [\Msun] & $ 0.73 \pm 0.1   $ \\
    Stellar radius $ \mathrm{R}_* $ [\Rsun]  & $   0.678^{+0.028}_{-0.032}$\\
    &    &      &  \\
   Planet mass  $ \mathrm{M}_p $ [\Mjup]  & $ 3.14 \pm 0.27 $\\
   Planet radius $ \mathrm{R}_p $ [\Rjup]  &  $1.039^{+0.043}_{-0.049}$ \\
   Planet density  $ \rho_p $ [$\rho_J$]  &  $2.81^{+0.44}_{-0.33}$ \\
    \hline
  \end{tabular}
\end{table}

  \section{Discussion} 

\subsection{Additional perturber}
\label{masslimit}

From the analysis of the TTVs, \citet{Maciejewski2011}
postulated the presence of a perturber with a mass of $\sim 0.1$ $M_{\rm
Jup}$ on a circular orbit with a period of $\sim 5.28$ days. This
orbital period places the system close a 5:3 mean-motion resonance. The
existence of the planetary companion was motivated by a large reduced
chi-square $\chi_{\rm red}^2=13.8$ associated to a linear fit of the mid-transit times.
 
\begin{figure}
\begin{center}
\includegraphics[width=\linewidth,viewport=0 0 435 165,clip]{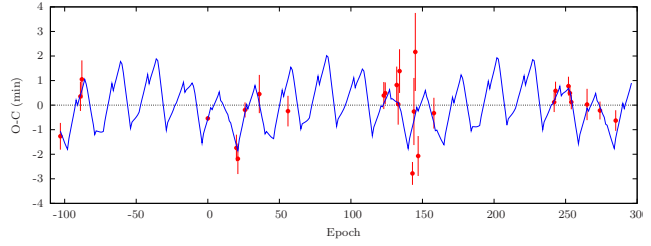}
\caption{\label{solution} Comparison between numerical TTVs (solid
blue curve) produced by an hypothetical perturber and the timing residuals (red dots
with error bars).}
\end{center}
\end{figure}

In the present study, additional transit observations and new
estimations of the error bars lead to $\chi_{\rm
red}^2=3.1$ for a linear ephemeris. This value is lower than the one obtained by
\citet{Maciejewski2011}, but remains significantly larger
than one, suggesting the existence of an additional signal. As in
\citet{Maciejewski2011}, we make the hypothesis that this
signal is produced by a planetary companion and try to constrain its
mass and orbital properties. For that purpose, we performed $24\times
10^5$ numerical 3-body simulations with 400 values of the companion
orbital period $P_c$ ranging uniformly between 1.5 $P_b$ to 10 $P_b$, 60
values of the initial longitude $\lambda_c$ (at the time of the first
transit Epoch=-103) uniformly distributed over 360 degrees, and 100
masses $M_c$ between 0.01 to 5 Jupiter masses with logarithmic steps. In
all simulations, the initial conditions correspond to circular coplanar
orbits. The best fit to the timing residuals is very isolated in the parameter space
and gives $P_c=19.3942$ days, $M_c=0.2457$ $M_{\rm Jup}$, and
$\lambda_c=48$ deg with $\chi_{\rm red}^2=1.77$. 
The TTV signal produced by this set of parameters
is superimposed to the timing residuals in Fig.~\ref{solution}.

\begin{figure}
\begin{center}
\includegraphics[width=\linewidth,viewport=0 0 215 200,clip]{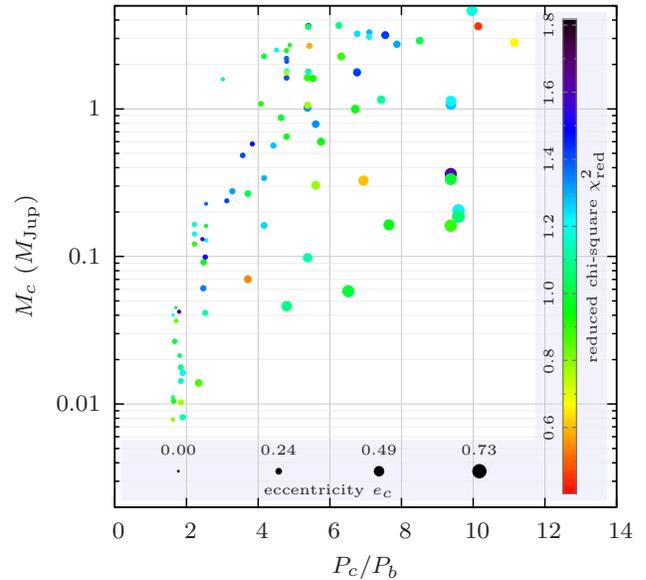}
\caption{\label{fig.ptmeth} Possible parameters of an hypothetical
planetary companion based on the timing residuals. Each point is the best
fit to a mock data set.}
\end{center}
\end{figure}

Since our best solution is completely different from that of
\citet{Maciejewski2011}, we analyse its robustness to the
 mid-transit time uncertainties. We generate 100 sets of artificial timing residuals
with the same 27 Epochs as the observations. The value of each point
is then taken from a normal distribution whose mean corresponds to our
best solution and width equals the uncertainty of that Epoch.
To reduce the cpu time, we forgo the 3-body simulations in favour of 
the perturbation method {\tt ptmeth} \citep{Nesvorny2008}.
This is a powerful analytical method about $10^4$ faster than numerical
simulations which provides accurate results outside of mean-motion
resonances \citep{Nesvorny2008}. It is also the
technique used by \citet{Maciejewski2011} to obtain a good
estimate of the location of the best parameters before performing
numerical simulations. The best solutions derived for each of the 100
artificial data sets are plotted in Fig.~\ref{fig.ptmeth}. Each point
correspond to a different data set. The most remarkable result is that
these best solutions, with reduced chi-square\footnote{The values of the reduced chi-square obtained with ptmeth are statistically lower than 1.77 found with the 3-body simulations. This is expected since the model of the perturbation method contains two additional parameters (eccentricity and longitude of periastron) in comparison to the slowest $N$-body integration method. Moreover, the mock data sets fitted by ptmeth 
are not affected by stellar activity.}  of the
order of or even lower than 1, are widely spread in the mass/period
diagram with $P_c$ between 1.5 and 12 $P_b$, and $M_c$ between 0.01 and 5
$M_{\rm Jup}$.  It is thus not currently possible to infer any set of
orbital parameters from the analysis of the timing residuals of WASP-10. This would
require lower error bars and/or a larger number of observations.

 It can be noted that the most massive companions at a given orbital period, in Fig.~\ref{fig.ptmeth}, are those with the lowest eccentricity. This is expected since eccentricities increase the orbital interactions and hence the TTVs. This property has already been used on different systems to set an upper boundary on the mass of a perturber as a function of its semi-major axis \citep[e.g.,][]{Steffen2005} . For instance, in the case of WASP-10, from the $N$-body simulations we obtain a 3 sigma upper mass equal to $\sim 0.1 M_{\rm Jup}$ at the 5.28 day period claimed by  \citet{Maciejewski2011}. This is not stringent enough to rule out the solution of  \citet{Maciejewski2011}, and neither does it confirm it since it lies at the 3 sigma limit and not at the best solution. Moreover, it should be stressed that any perturber with mass beyond 
$\sim 1$ $M_{\rm Jup}$ would have been detected by radial velocities.
They can thus be discarded.

This analysis illustrates the difficulties of interpreting the perturbations produced by a non-transiting companion in very low signal to noise data \citep[see, e.g.,][]{Boue2012}. The detection and the characterisation of non-transiting planets do require strong TTV signals as in the cases of Kepler-19 and KOI-872 \citep{Ballard2011, Nesvorny2012}.

\subsection{The effect of spots in measured TTVs}

Spot occultation features have been shown to affect measured transit times \citep{MillerRicci2008,Alonso2009,SanchisOjeda2011}.
The transit light curves of WASP-10b show features that might be due to the occultation of stellar spots by the planet.
\citet{Maciejewski2011b} shows that the features in epochs 242, 243, 253 are compatible with spots along the transit cord 
with an area of approximately 25\% -37\% that of planetary disc. However, the same shapes features can also be caused by systematic noise \citep{Barros2011b}.
In this section,  we will test the effect of spot features in the light curve shape and consequently in the fitted parameters, particularly in the measured transit times.
Since we are analysing data from different instruments, filters, and signal-to-noise ratio we will not attempt to test if the "spot features" 
are due to real spots occultations or systematic noise.
Instead, we assume that independently of its causes, a "spot feature" will have similar effect on the shape of the light curve.

\begin{figure}
  \centering
 \includegraphics[width=\columnwidth]{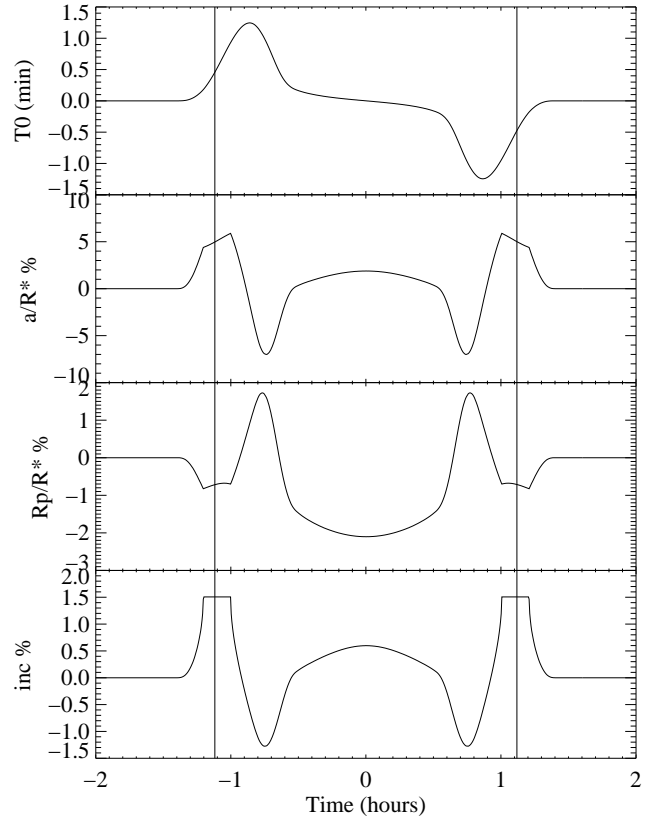}
  \caption{The figure shows the effects of a spot in the fitted light curve parameters as a function of mid spot time relative to mid transit time. The vertical lines show the position of ingress and egress.}
  \label{spotlc}
\end{figure}

\begin{figure}
  \centering
 \includegraphics[width=\columnwidth]{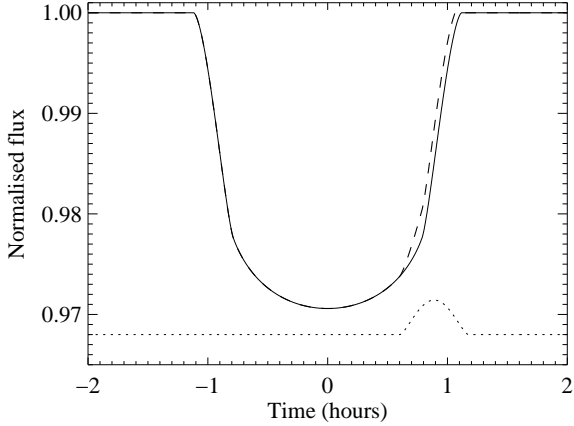}
  \caption{We show a transit light curve with an occulted spot during egress as a dashed line. It results from a combination of a transit light curve model shown as a solid line and a spot feature shown as a dotted line. Although there is no evidence of a spot in the combined light curve, it has a shorter duration and a measurable TTV=$-73$ seconds.}
  \label{spotegress}
\end{figure}

We modelled the spot features as inverted transits, a procedure similar to \citet{Kundurthy2011} and \citet{Maciejewski2011b}. 
We used the transit model described above, but for the spot features we assume no limb darkening and zero planet-spot impact parameter. 
Values for $ a $ and  $ \mathrm{R}_p $ were fixed to those of Table~\ref{MCMC} and an extra factor was included to account for the spot-to-star contrast.
This is a very simplified spot model, but reproduces the observed features quite well.
To obtain typical values for the spot parameters for our simulations we used the spot seen in light curve epoch 265 as a template. 
According to our tests of light curve systematics this appears to be a real spot.

To simulate the effects of a spot in a transit light curve we added the model spot to the model light curve of WASP-10b at different in-transit phases. 
The resulting light curve was then fitted with a transit model not including a spot to test how the spot affects the fitted parameters. 
Errors were not included in the simulated light curves. 
 We also do not account for geometrical foreshortening \citep{SanchisOjeda2011}, i.e. the same spot amplitude was used for each phase. 
The derived parameters as a function of central spot time relative to the mid-transit time are shown in Figure~\ref{spotlc}, where evidently a spot similar to the one in light curve epoch 265 is able to create fake transit time variations up to 1.2 minutes. 
Clearly, the spots will also significantly affect the derived shape parameters: $a/R_*$, $ R_p/R_* $,  and the inclination.
For the best signal-to-noise light curves, the spot effect will be more than $3\, \sigma$ of the errors quoted in Table~\ref{resm1}.

Noteworthily, the effect of the spot is higher when the spot is in the limb of the star,
 although spots at longitudes up to 30 degrees from the limb will significantly affect the measured transit times.
Furthermore, spot features during ingress and egress are harder to detect in the light curve both analytically and by eye.
%we note that when the TTV effect of the spot is higher, during ingress or egress, its presence in the light curve 
This is illustrated in Figure~\ref{spotegress} which shows an example of a light curve where the planet occults a spot during egress.
 The combined light curve is well fitted by a transit model with a shorter duration and earlier egress time which results in an earlier T0.
Moreover, in individual fits this light curve would appear to have low red noise.

\subsection{Spot modelling}
As seen in the previous section, the light curves that show higher TTVs due to spots also show a larger $a/R_*$
 (or shorter duration). Hence, a high value  $a/R_*$  is an indication that the light curve could be affected by spots. 
 Moreover, in this case the red noise might be underestimated for individual fits.
To check if the TTVs presented in Table~\ref{resm2} are affected by spots we tested the light curves that have 
a high red noise and also the ones with high $a/R_*$.

 From the results in Table~\ref{resm1}, as mentioned before, we conclude that two light curves have higher $a/R_*$ than the mean value. 
 These are epochs 20 and 147.
Table~\ref{beta} shows that five light curves have red noise higher than 2 for model 2, namely epochs 143, 145, 147, 254b, and 265.
These 6 light curves were fitted with a transit model including a spot model as described in the previous section. The impact parameter was fixed to 0 and we assume no limb darkening,  the contrast, phase, and duration of the spot were fitted. When we include a spot feature we find variations on the fitted transit times up to 2.2 minutes (for epoch 20).
We extended the analysis to all the light curves, but found T0 variations only in another light curve; epoch 21.
The difference between the derived T0 when including a spot model and the T0 derived with model 2 are given in 
Table~\ref{tspot} as well as the corresponding red noise for the one spot model parametrised by $\beta$.
  For curiosity, we also give the spot fitted parameters in Table~\ref{tspot}, however the fits are highly degenerate and dependent on the starting parameters.
 We confirm that for all the light curves, the time correlated features can be well described by a spot, given that the estimate of the red noise significantly decreases.
 However, in some cases such as epoch 147 residual red noise remains in the light curve after accounting for one spot.
Although the fits are highly degenerate, we confirm that only spots close to the limb of the star have a significant effect on the measured transit times.  Although the spot induced TTV can be large, the new transit times agree with model 2 results.  However, this would not be the case if we had not included red noise. Furthermore, the existence of a spot explains the significant TTVs found for model 1 which does not penalise shape variations. Hence, in this case model 1 underestimates the errors in the mid-transit times.

\begin{figure*}
  \centering
  \includegraphics[width=1.8\columnwidth]{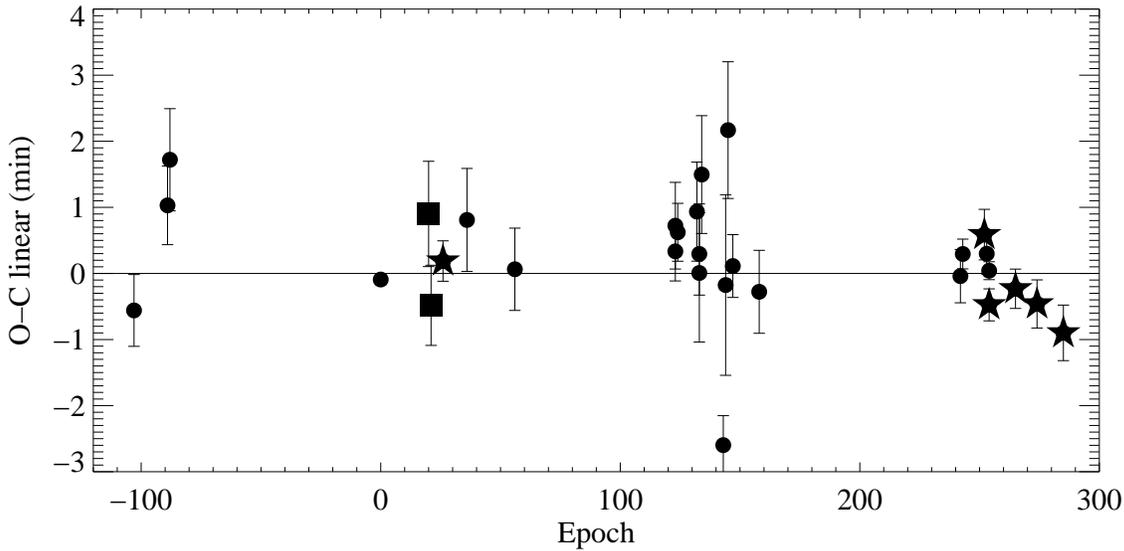}
  \caption{ Spot model transit residuals of WASP-10 after subtracting a linear ephemeris.
   The symbols are the same as Figure~\ref{timeplot}. RISE data are shown as filled stars, FTN data as squares while preciously published data as circles.}
  \label{timeplotspot}
\end{figure*}

 In Figure~\ref{timeplotspot}, we show the timing residuals for the one spot model.  The  $\chi^2$ of a linear ephemeris is 76 for 28 degrees of freedom and the p-value is  $1.6 \times 10^{-6}$ which would lead to a rejection of the model. Epoch 143 is still inconsistent ($5.8\,\sigma$) with a linear ephemeris since this discrepancy cannot be explained with a spot model.
 The ingress of this transit is clearly earlier than predicted by the linear ephemeris which cannot be caused by a spot.  If we exclude epoch 143 the $\chi^2$ of the linear ephemeris reduces to 43 for 27 degrees of freedom ($\chi_{\rm red}^2$= 2.56) and the p-value is 0.028 and hence a linear ephemeris is no longer rejected. Therefore, we conclude that if we account for ``spot features'' the transit times of WASP-10 are consistent with a linear ephemeris.
%This implies that the spot feactures in transit 20, 21 and 147 are the cause of the $\chi^2$  of a linear fit to the model 2 derived transit times

\begin{figure}
  \centering
  \includegraphics[width=\columnwidth]{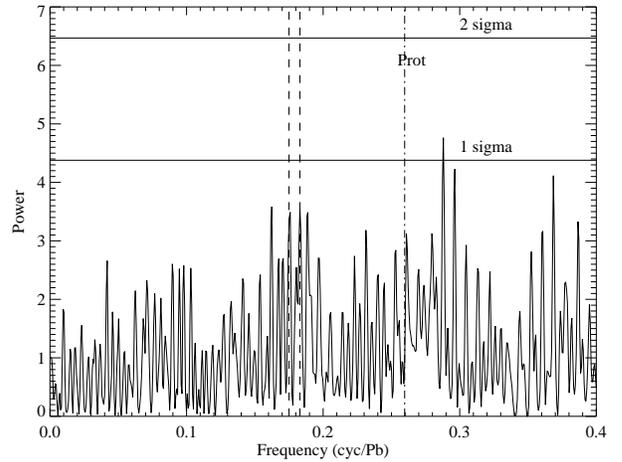}
  \caption{Lomb-Scargle periodogram of residual transit times from a spot model after subtracting a linear ephemeris. The vertical dash-dotted line shows the rotation period of 11.91 the dashed lines show the TTV frequencies proposed by \citet{Maciejewski2011} and the horizontal lines shows the false alarm probability at 1 and $2\, \sigma$ confidence limit. }
  \label{peri_plot}
\end{figure}

Figure~\ref{peri_plot} shows the Lomb-Scargle periodogram of the residuals from a linear ephemeris. Notice that the peaks close to the rotation period of the star have decreased.   The main difference between this figure and the results for model 2 is due to the ``spot features'' during egress found for epochs 20, 21 and 147. Given that these ``spot features'' add power at the rotation period of the star, this suggests that these spots are real rather than systematic noise.

However, we caution the reader to an over interpretation of these results because our model is very simplified, the fits are degenerate, and we cannot confirm that all the features present in the light curve are real spots. 
A complete spot model of WASP-10 is beyond the scope of this paper, our objective being to estimate the mid-transit timing errors that can be caused by spots.

In summary, stellar spots or correlated noise can significantly bias the derived parameters of the light curves especially if the spot occurs during ingress or egress.
 However, the errors due to spots can be accounted for if we include red noise due to shape variations as done for model 2.

  \begin{table}
  \centering 
  \caption{Variations of the fitted mid-transit times for the one spot model. We also show the estimated red noise of the resulting residuals and the spot parameters.}
  \label{tspot}
  \begin{tabular}{lccccc}
    \hline
    \hline
Epoch & delta T0 & $\beta$ & $T_{spot}-T0$ &heigh & $tdura$ \\
   & (sec) & &  (min) & (mmag) & (min) \\
    \hline
20    & -135  &   1. & 50.72 & 7.98 & 31.5\\
21    &  -78  &   1. & 36.97 & 3.13  & 62.58\\
143   &   -6  &   1.4& 1.47 &1.61 &46.16\\
145   &    5  &   1.3& -10.50 & 8.22 & 101.94\\
147   & -126  &   2. &  52.38 &5.93 &52.60\\
243   &    7  &   1.1& 39.73  & 3.03 &   -10.28\\
254b  &   17  &   1. &  -26.48 &1.97  &55.69 \\
265   &    2  &   1. &  32.35 & 3.99 &28.71\\
  \hline
  \end{tabular}
\end{table}

\subsection{Effects of partial transits}
\begin{figure}
  \centering
  \includegraphics[width=\columnwidth]{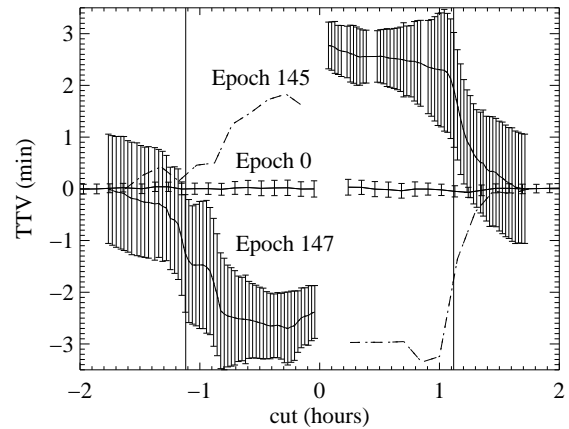}
  \caption{ Induced TTVs as a function of the time the light curves were cut, Tp. We show the errors accounting for red noise estimated for each fake partial light curve.
   The errors of epoch 145 were excluded for clarity. These errors are higher than for epoch 147, and hence, the induced TTVs are less significant. The vertical lines show the position of ingress and egress.}
    \label{partial}
\end{figure}

Several authors have noted that the uncertainties in the T0s derived from partial transit are underestimated  \citep[e.g][]{Gibson2009, Barros2012}. 
Given that the one transit that shows significant transit timing variations is a partial transit (epoch 143),
%Given that 9 out of the 30 light curves for WASP-10b are partial transits,
 in this section we illustrate how the derived transit times are affected in partial transits.
We assume here that the light curves are fitted simultaneously (model 2), and hence the shape of the transits is set by
 the complete transits, i.e. we fixed  $a/R_*$ and  $i$  in our test fits.
To test the TTVs induced by partial transits, we used some of our complete light curves, from which we subtract model 2 estimated T0 ( Table~\ref{resm2}).
The light curves were then gradually cut at a time -Tp or Tp which corresponds to cutting the beginning or end of the observation respectively.
Resulting partial light curves were fitted to derived the induced T0 variation. Red noise was estimated for each fake light curve and included in the errors. 
In Figure~\ref{partial} we show our test applied to the light curves epoch 0, 145, and 147 as examples. For other light curves the derived TTVs are enclosed in the examples presented. 
We conclude that light curves that do not have red noise such as epoch 0 show partial transit induced TTVs that are consistent with zero.
 However, for light curves with high red noise the induced TTVs can reach several minutes and the derived errors are indeed underestimated. 
 The errors decrease with Tp due to red noise underestimation. 
 For example, the TTVs derived from fake partials of epoch 147 are significant at $5\, \sigma$ even when we include red noise.
 This is caused by the correlation between red noise and the normalisation parameters $F_{out}$  and $T_{grad}$ in partial light curves.
  Notice that TTVs are induced even when Tp is larger than half of the transit duration, i.e. the transit needs to include some out of transit data.  Therefore this could explain the 2.7 minutes TTV found for epoch 143, though we remind that this is also the only transit taken with 2.0m Rozhen telescope.

%We attempted to reproduced the TTV measured for epoch 143 by simulating a partial light curve with a spot. 
%In this particular case, we can produce negative TTV by including a spot during egress. However, as already mentioned above a spot model does not fit the observed light curve well.

 If we discard all partial transits, the transit residuals of model 2 are consistent with a linear ephemeris, with a $\chi^2=30$ for 17 degrees of freedom ( $\chi_{\rm red}^2=1.8$) and a p-value $=0.025$. Model 1 results are still inconsistent with a linear ephemeris at the $5 \sigma$ level since transits show significant transit time variations due to spots that are not accounted for in this model as was shown in the previous section. The one spot model timing residuals of the complete transits are also consistent with a linear ephemeris: $\chi^2=20$ for 17 degrees of freedom ($\chi_{\rm red}^2=1.2$) and a p-value $=0.27$.  The increase of the p-value when we discard partial transits indicates that errors in transit times from partial transits are indeed underestimated, as discussed above.

In summary, it is possible to obtain a good estimation of the mid-transit times of partial transits if they do not have red noise.
However, in the presence of red noise the errors of the mid-transit times of partial transits can be significantly underestimated due to our inability to recognise red noise in partial light curves.
 Unfortunately, complete transits are sometimes unattainable due to weather and technical issues. 
For example, all scheduled  RISE observations were intended to cover the complete transit.
We conclude that transit times derived from partial transits have to be regarded with extreme care.

\section{Conclusion}
We have presented eight new transit observations of WASP-10b with RISE mounted on the Liverpool telescope and Merope mounted on the Faulkes Telescope North.
The new observations were combined with 22 previously published transit light curves to derive the transit times homogeneously.
System parameters were re-derived assuming a circular orbit and assuming that the star is less active at maximum brightness.
However, due to bright spots some stars, like the Sun, are actually brighter at the activity maximum. 
No bright spot is seen in the transit cord of WASP-10, therefore, we assume that the observed stellar modulation is due to dark spots.

  We show that the significant transit timing variations for WASP-10b reported by \citet{Maciejewski2011,Maciejewski2011b} were due to the normalisation parameters being neglected in some of previous fits to the data.
 We find a time difference of 120 and 49 seconds for epochs 36 and 147, respectively when we account for a baseline function in the transit fit. It is crucial to account for the baseline function because it correlates strongly with the estimated transit times.

 We present two models to derive the transit times. Model 1 where the transits are fitted individually and model 2 where the transits are fitted simultaneously. Comparison between the two models lead us to suspect that the transits times were affected by spot features or systematic noise.
We show that spot features or systematic noise during ingress and egress have a significant effect on the measured transit times leading to spurious TTV detections. Since they also cause transit duration variations they are treated as red noise in model 2 which explains the differences found between the two models. Nevertheless, the transit times derived with model 2 show higher dispersion than expected for a linear ephemeris.

We tested if the transit times could be explained by a perturber model. As expected the addition of a planetary companion in the system reduces the rms of the residuals. However, there is not more evidence for the TTV signal claimed by \citet{Maciejewski2011,Maciejewski2011b} neither in the periodogram nor in the n-body analysis of the TTVs. In fact, we found that other configurations with perturber masses ranging from 0.01 to 5 \Mjup and period between 1.5 and $12\times$ the orbital period of WASP-10b fit the data equivalently well. Therefore, we conclude that the transit times of WASP-10b have insufficient accuracy and precision to validate the presence of any planetary companion.

 An alternative model for the dispersion seen in the transit times of WASP-10b is stellar activity induced TTVs. We tested this hypothesis by including a spot feature in the transit model fit. Our simple model shows that transits 20, 21 and 147 are consistent with having a spot feature during egress that results in transit time variations of -135, -78 and -126 seconds respectively.
 This explains the dispersion found for the transit times derived with model 2. We suggest that this spot induce transit variations are due to real spots rather than systematics because they add power to the residuals at periods close to the orbital period of the host star. Moreover, other spots are seen during transit, in particular epochs 242, 243 and 253 which were studied in detail by \citet{Maciejewski2011b} that concluded that they were consistent with real spots.
 When we account for these spot features, the derived transit times are consistent with a linear ephemeris with the exception of epoch 143, which we show it could be due to this transit being partial.

 We also show that partial transits may produce fake TTVs up to 3 minutes, and hence T0s derived from partial transits should be regarded with extreme care. When we discard partial transits the dispersion of the transit times relative to a linear ephemeris decreases significantly both for model 2 results and for the one spot model results.

\section{Acknowledgements}
We thank Damian Christian, John Johnson, Jason Dittmann, Tereza Krejcova, and Gracjan Maciejewski for kindly sharing their light curves on WASP-10 and for interesting discussions. We thank the anonymous referee for contributing to the improvement of this manuscript. We thank Tom Marsh for the use of the
ULTRACAM pipeline.
SCCB thanks John Southworth and Rodrigo D\'iaz for interesting discussions.
 The RISE instrument mounted at the Liverpool Telescope was
designed and built with resources made available from Queen's
University Belfast, Liverpool John Moores University and the
University of Manchester. The Liverpool Telescope is operated on the
island of La Palma by Liverpool John Moores University in the Spanish
Observatorio del Roque de los Muchachos of the Instituto de
Astrofisica de Canarias with financial support from the UK Science and
Technology Facilities Council.

\bibliographystyle{mn2e} 

\bibliography{susana_mn}

\label{lastpage}

\end{document}